\documentclass[a4paper,11pt]{article}
\usepackage{jheppub} 
\usepackage{amssymb,amsmath,xcolor,empheq,physics,mathtools,url}

\numberwithin{equation}{section}
\usepackage{epsfig}
\usepackage{epstopdf}
\usepackage{latexsym}
\usepackage{graphicx}
\usepackage{booktabs}
\usepackage{bbm}
\usepackage{enumitem}
\usepackage{subcaption}

\usepackage[toc]{appendix}
\usepackage{genyoungtabtikz}
\usepackage{ytableau}
\usepackage{caption}    
\usepackage{float}       
\usepackage{array}
\usepackage{booktabs}
\usepackage{multirow}
\usepackage[table]{xcolor}
\def\e{\exp}
\def\x{\mathbf{x}}
\def\y{\mathbf{y}}
\def\z{\mathbf{z}}

\def\ii{\mathrm{i}}

\def\normord{ {\scriptstyle {{\bullet}\atop{\bullet}}} }

\DeclareMathOperator{\powsym}{p}

\allowdisplaybreaks[1]
\arxivnumber{} 

\title{\boldmath Unitary matrix models, quantized symmetric functions and spin chain}
\preprint{
    USTC-ICTS/PCFT-26-41
}

\author[1,2]{Gao-fu Ren}
\affiliation[1]{Interdisciplinary Center for Theoretical Study, University of Science and Technology of China,\\Hefei, Anhui 230026, China}
\affiliation[2]{Peng Huanwu Center for Fundamental Theory, Hefei, Anhui 230026, China}
\emailAdd{rengaofu@mail.ustc.edu.cn}

\abstract{We construct a correspondence between a broad class of unitary matrix models and vacuum correlation functions in quantum spin-chain Hilbert spaces. The key step is to lift symmetric functions to operators acting on the $N$-magnon sector in a way that preserves the relevant ring structure. For any unitary matrix model whose integrand admits a factorized expansion in symmetric functions, the Schur orthogonality of the unitary group integral is then translated into the inner product of quantized Schur states. We illustrate the construction for the Gross--Witten--Wadia model, superconformal indices of $\mathcal{N}=4$ super Yang--Mills theory with classical gauge groups, and Toda tau functions. The resulting operator formulation provides a unified algebraic bridge between unitary matrix integrals, quantum integrable systems and symmetric function theory.}

\begin{document}
\maketitle
\flushbottom
\section{Introduction}
A unitary matrix model is a matrix integral over the unitary group $U(N)$ equipped with the Haar measure, typically of the form
\begin{align}
    Z=\int_{U(N)}\mathcal{D}U\exp(\mathrm{Tr}V(U,U^\dagger))
\end{align}
where the potential $V$ is invariant under unitary conjugation. Originally introduced as a tractable setting for large $N$ gauge theories, unitary matrix models are now connected with gauge theory, quantum gravity, topological strings and integrable systems. More recently, the finite 
$N$ case has attracted considerable interest.

One of the most outstanding achievements of matrix models is that the double scaling limit of Hermitian matrix models provides a naturally regularized, non-perturbative definition of non-critical string theories, and reveals deep connections with classical integrable hierarchies, notably the KdV type hierarchy~\cite{Douglas:1989ve,Banks:1989df}. It is therefore natural to expect that the powerful techniques and structural insights developed for Hermitian matrix models carry over to unitary matrix models. The simplest and most extensively studied unitary matrix model is the Gross--Witten--Wadia (GWW) model~\cite{Gross:1980he,Wadia:1980cp,Wadia:2012fr} and its generalizations~\cite{Russo:2020eif}. Its main interest lies in the Gross--Witten transition. This transition has attracted considerable attention because it has been shown, through holographic duality and certain large 
$N$ limits, to capture essential features of the Hawking--Page phase transition of black holes and the confinement/deconfinement transition in QCD~\cite{Dumitru:2004gd,Copetti:2020dil}.

A second class of examples considered below is furnished by the superconformal index of $\mathcal{N}=4$ super Yang--Mills (SYM) theory. This index can be written as a unitary matrix integral~\cite{Kinney_2007} and can be evaluated explicitly using the method known as character expansion~\cite{Sei:2023fjk} via the orthogonality of Schur functions.
The connections between general superconformal indices and symmetric functions have been extensively studied \cite{Hatsuda:2025mvj,Ren:2025tvx,Ren:2026dlj,Gadde:2011uv,Gaiotto:2012xa,Razamat:2013qfa,Rastelli:2014jja}, revealing a profound relation to classical integrable systems. In the most general case, without taking any limit of the fugacities, the superconformal index can be decomposed into two parts. One part can be expressed in terms of elliptic Macdonald polynomials, which are the eigenstates of the elliptic Ruijsenaars--Schneider model on $\mathbb{T}^N$, and involves certain structure constants. The other part is an integral with the elliptic measure, which defines the orthogonality relations of the elliptic Macdonald polynomials.
The phase transition of the $\mathcal{N}=4$ SYM superconformal index with a truncation was investigated in~\cite{Dutta:2007ws}, and such a transition is expected to be dual to the Hawking--Page transition in the holographic dual.
Moreover, a wide class of tau functions of Toda type can also be represented as unitary matrix integrals~\cite{Martinec:1990qg}. In particular, the superconformal index of $\mathcal{N}=4$
$U(N)$ SYM is related to such a Toda tau function, up to a generalized Hubbard--Stratonovich transformation~\cite{murthy:2022ien}.

Our main focus is the relation between unitary matrix models and quantum integrable systems. We use the free-fermion construction together with quantized symmetric functions to reformulate unitary matrix integrals as spin-chain correlators. Spin chains and bosonic chains provide standard examples of quantum integrable systems, and their Hilbert spaces can be embedded naturally into a free-fermion Fock space; in this framework one has a one-to-one map between the relevant bases~\cite{Wheeler:2010vmq}.
In recent work~\cite{crichigno:2024aub}, the algebraic ring of symmetric functions was reformulated directly in the Hilbert space of spin chains, using the quantization formalism of~\cite{gelfand:1994hg,fomin1998179}. This reformulation is particularly relevant for us because it identifies the algebraic relation of classical symmetric functions with operators acting on the spin-chain Hilbert space. Building on these ideas, \cite{Bravyi:2025tum} introduced a classical algorithm for computing the characters of the symmetric group using an MPS--MPO (matrix product state--matrix product operator) construction. This algorithm becomes significantly more efficient than the standard approach based on the Murnaghan--Nakayama rule when the Young diagrams involved are long. Quantum algorithms and quantum circuits for characters of symmetric groups were also developed in~\cite{beals1997quantum,kawano:2015ozw}, and explicit quantum circuits for calculating characters are given in~\cite{Bravyi:2025tum}.
In addition, the connection between black holes and spin chains has attracted considerable interest in recent years~\cite{Kristjansen:2025xqo}. The accessibility of the Hawking--Page transition in AdS$_5$ through a one-dimensional Heisenberg spin chain
is demonstrated in~\cite{Perez-Garcia:2024pcq}.

The main result of this work is that a broad class of unitary matrix models can be recast as vacuum correlation functions in the $N$-magnon sector of a spin-chain Hilbert space. The construction relies on a quantization of symmetric functions that preserves the ring structure relevant to Schur expansions. After diagonalizing the unitary matrix, the original integral is reduced to an eigenvalue integral whose integrand can be expanded in Schur functions. The Schur orthogonality relations are then translated into inner products in the spin-chain Hilbert space. The quantization procedure is introduced in Section~\ref{Classical and quantized symmetric function theory}, and the matrix model correspondence is derived in Section~\ref{Unitary matrix model and quantized symmetric function}.
We illustrate the framework with the GWW model, superconformal indices of $\mathcal{N}=4$ SYM theories and Toda tau functions.

The paper is organized as follows.
In Section~\ref{Heisenberg spin chains and Free fermions}, we introduce the Hilbert space of spin chains and explain how it can be constructed using the free-fermion formalism~\cite{Wheeler:2010vmq}.
In Section~\ref{Classical and quantized symmetric function theory}, we present the guiding philosophy for quantizing symmetric functions: roughly speaking, the classical ring of symmetric functions is lifted to an algebra of operators acting on the basis in spin-chain Hilbert space, with its algebraic structure preserved. We also collect the essential mathematical facts that will be used in our paper.
In Section~\ref{Unitary matrix model and quantized symmetric function}, we derive our main result: a vast class of unitary matrix models can be recast as a vacuum correlation function of an operator built from quantized symmetric functions, evaluated in the $N$ magnon sector of the spin-chain Hilbert space. We illustrate this general construction with several concrete examples, including the Gross--Witten--Wadia model and the superconformal index of $\mathcal{N}=4$ super Yang--Mills theory.
We conclude in Section~\ref{Conclusion} by summarizing our results and outlining several promising directions for future research.

\section{Heisenberg spin chains and free fermions}\label{Heisenberg spin chains and Free fermions}
In this section we collect the spin-chain and free-fermion ingredients used in the rest of the paper. The purpose is twofold. First, we recall how the fixed-magnon sectors of a spin chain are naturally indexed by Young diagrams. Second, we explain how the same data arise from charged free-fermion Fock spaces after imposing suitable projection conditions. This provides the Hilbert-space dictionary on which the quantized symmetric functions of the next section will act.

\subsection{Heisenberg spin chains}
\label{Heisenberg spin chains}
We begin with a spin-$\frac{1}{2}$ chain with $n$ sites. Its Hilbert space is
$\mathcal{H}=(\mathbb{C}^2)^{\otimes n}$, and at each site we use the local basis
$(\ket{\uparrow},\ket{\downarrow})$. The general nearest-neighbour Heisenberg Hamiltonian is
\begin{align}
    \label{XYZHamiltonian}
H_{\text{XYZ}}= \sum_{i}( J_{x} \, \sigma^x_{i+1}\sigma^x_i + J_y \, \sigma^y_{i+1}\sigma^y_i + J_z \, \sigma^z_{i+1}\sigma^z_i)\,,
\end{align}
where $\sigma^x_i,\sigma^y_i,\sigma^z_i$ are the standard Pauli matrices acting on the spin state at site $i$
\begin{align}
    \sigma^x_i=\begin{pmatrix}
    0 & 1 \\
    1 & 0
  \end{pmatrix}_i\,,\qquad\sigma^y_i=\begin{pmatrix}
    0 & -\ii \\
    \ii & 0
  \end{pmatrix}_i\,,\qquad\sigma^z_i=\begin{pmatrix}
    1 & 0 \\
    0 & -1
  \end{pmatrix}_i\,.
\end{align}
Here $J_{x,y,z}$ are real coupling constants, and the interaction is restricted to nearest neighbor sites. The most general Heisenberg spin chain for which $J_{x,y,z}$ take arbitrary values is known as the XYZ quantum spin chain. Important special cases include:
\begin{enumerate} 
\item $J_x=J_y=0$, $J_z\neq0$. Ising spin chain.
\item $J_x=J_y\neq0$, $J_z=0$. XX spin chain, which is equivalent to a free lattice 
fermion by Jordan-Wigner transformation.
\item $J_x=J_y=J_z\neq0$. XXX spin chain.
\item $J_x=J_y\neq J_z\neq0$. XXZ spin chain.
\item $J_x\neq J_y\neq J_z\neq0$. XYZ spin 
chain.
\end{enumerate}

The Bethe ansatz gives a useful description of eigenstates in these integrable spin chains. We focus on the XXZ case, $J_x=J_y=J$, $J_z=\Delta$, because its fixed-magnon sectors will be the ones used below. The total Hilbert space decomposes as
$\mathcal{H}=\oplus_{i=0}^\infty\mathcal{H}_i$, where $\mathcal{H}_i$ is spanned by states with exactly $i$ down spins. In the $k$-magnon sector, a basis vector is labelled by the ordered positions of the down spins,
$\ket{n_1,\ldots,n_k}$, and the Hamiltonian preserves this grading.
In this sector the coordinate Bethe ansatz writes an eigenstate as

In the $k$ magnon sector the Bethe ansatz represents an energy eigenstate as a superposition of such basis vectors,
\begin{align}\label{XXZCBA}
\ket{\Psi(\mathbf{p})} = \sum_{1 \le n_1 < \cdots < n_k \le n} \Psi(n_1,\dots,n_k) , \ket{n_1,\dots,n_k},
\end{align}
with wavefunction
\begin{align}
\Psi(n_1,\dots,n_k) = \sum_{\sigma \in S_k} \exp\Bigl( \ii\sum_{a=1}^k p_{\sigma_a} n_a + \frac{\ii}{2}\sum_{a\neq b} \Theta(p_{\sigma_a},p_{\sigma_b}) \Bigr).
\end{align}
The phase $\Theta(p,q)$ encodes two-particle scattering, and we denote the corresponding scattering amplitude by
$S(p,q):=\mathrm{e}^{\ii\Theta(p,q)}$. Periodic boundary conditions impose the Bethe equations
\begin{align}\label{betheeqs}
\mathrm{e}^{\ii p_a n} = \prod_{b \neq a} S(p_a, p_b), \qquad a = 1,\dots,k.
\end{align}
Solutions $\{p_a\}$ of these equations are called Bethe roots. The remarkable consequence of the Bethe equations is that the many-body wavefunction is completely fixed by the two-body scattering matrix, a feature of quantum integrability.
For the XXZ chain, the scattering amplitude takes a particularly simple form when expressed in terms of the rapidities $u_a = 2 \cot \frac{p_a}{2}$
\begin{align}
S(p_a,p_b) = \frac{u_a - u_b + \ii\Delta}{u_a - u_b - \ii\Delta}.
\end{align}
In the isotropic limit $\Delta=0$, corresponding to the XX model, the scattering matrix becomes unity, $S(p_a,p_b)=1$. The Bethe equations then reduce to $\mathrm{e}^{\ii p_i n}=1$, implying that the variables $x_a:=e^{\ii p_a}$ are $k$ distinct $n$-th roots of unity, considered up to permutation. The space of Bethe roots is thus the configuration space
\begin{align}\label{XXBetheroots}
\mathcal{V}_{k,n} = \{ (x_1,\dots,x_k) \mid x_a^{n}=1,x_a \neq x_b \} / S_k.
\end{align}
Its dimension is $\text{dim}(\mathcal{V}_{k,n} )={n \choose k}$ , which precisely matches the dimension of the $k$ magnon sector. Moreover, the Bethe vectors \eqref{XXZCBA} evaluated at these roots constitute an orthogonal basis of $\mathcal{H}_k$.

\begin{figure}[t]
\begin{center}
\includegraphics{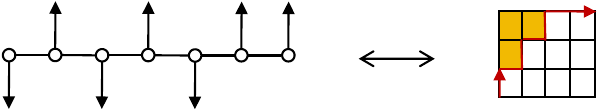}
\caption{$\ket{\downarrow\uparrow\downarrow\uparrow\downarrow\uparrow\uparrow}$ corresponds to Young diagram $(2,1)$}
\label{chainyoungcorrespondence}
\end{center}
\end{figure}

The Hilbert space of a one-dimensional spin chain can also be described using Young tableaux. Consider a spin chain of length $k$ in $N$ magnon sector (i.e., with exactly $N$ down spins). Any basis state is specified by assigning a spin  $(\ket{\uparrow}$ or $\ket{\downarrow})$  to each site, with the total number of down spins fixed to $N$. There is a well-known procedure that associates to each such state a Young diagram. One first draws a rectangle of size $N\times (k-N)$. Starting from the bottom-left corner, the sequence of spins defines a lattice path through this rectangle: reading the spins from site $1$ to $k$,  each down spin corresponds to a step upwards, while each up spin corresponds to a step to the right. The path thus runs from the bottom-left corner to the upper-right corner of the rectangle, never crossing its boundaries. The region enclosed by this path and the upper and left boundaries of the rectangle forms a Young diagram, which fits inside the $N\times (k-N)$ rectangle. The length of this Young diagram is not larger than $N$ and $\lambda_1\leq k-N$, for example, see Figure~\ref{chainyoungcorrespondence}. The correspondence between basis states and lattice paths is one-to-one, and different paths give rise to distinct Young diagrams. The number of such Young diagrams is precisely $\text{dim}(\mathcal{V}_{N,k} )={k \choose N}$, which equals the dimension of the $N$ magnon sector of the quantum XX spin chain. As will be shown below, these basis states are isomorphic to the basis states appearing in the free-fermion construction

According to this correspondence, operations on the quantum Hilbert space of spin chains translate into operations on Young diagrams. We will show that there is a formalism in which classical combinatorial properties can be lifted to the quantum level, giving rise to the theory of quantized symmetric functions. The vacuum of the $N$-magnon sector corresponds to the empty Young diagram:
\begin{align}
    \label{Nmagnonsectorvacuum}
    \ket{\emptyset,N}=\ket{\downarrow^N\uparrow\cdots}\,,
\end{align}
where $\downarrow^N$ denote $N$ down spins.

\subsection{Free fermions construction}
\label{Free fermions}
We now recall the free fermion Fock space construction and explain how the $N$-magnon sector is obtained from it. Let $\psi_n$, $\psi_n^*$, $n\in\mathbb{Z}$, be fermionic modes obeying
\begin{align}
    \{\psi_n,\psi_m\}=\{\psi^*_n,\psi^*_m\}=0\,,\qquad \{\psi_n,\psi^*_m\}=\delta_{nm}\,.
\end{align}
They generate an infinite dimensional Clifford algebra. It is convenient to package them into the free fermion fields
\begin{align}
    \psi(z)=\sum_{k\in\mathbb{Z}}\psi_kz^{-k}\,,\qquad\psi^*(z)=\sum_{k\in\mathbb{Z}}\psi^*_kz^{k}\,.
\end{align}
The vacuum state $\ket{0}$ is defined by  
\begin{align}
    \psi^*_n\ket{0}=0\,,\quad n<0\,; \qquad \psi_n\ket{0}=0\,,\quad n\geq0\,.
\end{align}
Similarly, the dual vacuum $\bra{0}$ satisfies
\begin{align}
    \bra{0}\psi_n=0\,,\quad n<0\,;\qquad \bra{0}\psi^*_n=0\,,\quad n\geq0\,.
\end{align}
This vacuum can be thought of as an infinite chain whose sites are labeled by the index $n$ of fermionic operators, with all sites $n<0$ occupied by a particle and all sites $n\geq0$ empty which can be thought of as a hole, in analogy with the `` Dirac sea " picture. A particle carries charge $-1$ and a hole carries the charge $+1$, hence any state with a definite number of particles and holes has a well defined total charge.

From this vacuum we can construct  `` shifted '' Dirac vacua $\ket{n}$ and their duals $\bra{n}$ 
\begin{align}
        \ket{n} =\left \{
     \begin{aligned}
     &\psi^*_{n-1}\cdots\psi^*_1\psi^*_0\ket{0}\,,\ && n>0\\
     &\psi_{n}\cdots\psi_{-2}\psi_{-1}\ket{0}\,,\ && n<0\\
\end{aligned}\right.\\\notag\\
\bra{n} =\left \{
\begin{aligned}
     &\bra{0}\psi_{0}\psi_{1}\cdots\psi_{n-1}\,,\ && n>0\\
     &\bra{0}\psi^*_{-1}\psi^*_{-2}\cdots\psi^*_n\,,\ && n<0\\
\end{aligned}\right.
\end{align}
The vacuum $\ket{n}$ carries charge $n$ and corresponds to the shifted Dirac sea in which all sites with $i<n$ are occupied by particles and all sites $i \geq n$ are empty.
 A convenient basis of the fermionic Fock space $\mathcal{H}_F$ with definite charge $n$ is provided by states $\ket{\lambda,n}$ labelled by Young diagrams $\lambda$. For a given Young diagram $\lambda=(\lambda_1,\cdots,\lambda_{\ell(\lambda)})$ with $\ell(\lambda)$ rows, let $(\Vec{\alpha}|\Vec{\beta})=(\alpha_1,\cdots,\alpha_{d(\lambda)}|\beta_1,\cdots,\beta_{d(\lambda)})$ is its Frobenius notation. $d(\lambda)$ is the number of boxes on the main diagonal, $\alpha_i=\lambda_i-i$, $\beta_i=\lambda_i'-i$, where $\lambda'$ is the transposed diagram of $\lambda$. Then the basis vectors are defined as
\begin{align}\label{deffreeferfocklambda}
    \begin{split}
        \ket{\lambda,n}:=\psi_{n-\beta_1-1}\cdots\psi_{n-\beta_{d(\lambda)}-1} \psi^*_{n+\alpha_{d(\lambda)}}\cdots\psi^*_{n+\alpha_1}\ket{n}\,,\\
        \bra{\lambda,n}:=\bra{n}\psi_{n+\alpha_1}\cdots\psi_{n+\alpha_{d(\lambda)}} \psi^*_{n-\beta_{d(\lambda)}-1}\cdots\psi^*_{n-\beta_1-1}\,.
    \end{split}
\end{align}
These states have total charge $n$, and the correspondence between Young diagrams and fermionic Fock states is one-to-one.  The bare vacuum $\ket{\infty}$,  the completely empty state is sometimes also considered, but it will not play a major role here. 

In order to isolate the subspace relevant to the $N$ magnon sector of the spin chain, we introduce projection operators that enforce certain occupation conditions. Following~\cite{alexandrov:2012tr}, we define projection operators which are also the non-invertible group-like elements
\begin{align}\label{projectionoperator}
\begin{split}
        &\mathrm{P}_n^+=\normord \e\left(\sum_{i< n}\psi_i\psi^*_i\right)\normord=\prod_{i< n}(1-n_i)=\prod_{i< n}\psi_i\psi^*_i\,,\\
         &\mathrm{P}_n^-=\normord \e\left(-\sum_{i\geq n}\psi_i\psi^*_i\right)\normord=\prod_{i\geq n}(1-n_i)=\prod_{i\geq n}\psi_i\psi^*_i\,,\\
         &\mathrm{P'}_n^+=\normord \e\left(\sum_{i< n}\psi^*_i\psi_i\right)\normord=\prod_{i< n}n_i=\prod_{i< n}\psi^*_i\psi_i\,,\\
         &\mathrm{P'}_n^-=\normord \e\left(-\sum_{i\geq n}\psi^*_i\psi_i\right)\normord=\prod_{i\geq n}n_i=\prod_{i\geq n}\psi^*_i\psi_i\,.
\end{split}
\end{align}
Here $n_i$ is the number operator at site $i$, it takes the value $0$ if the site is empty and $1$ if it is occupied.  
These operators satisfy the projector properties $(\mathrm{P}_n^\pm)^2=\mathrm{P}_n^\pm$ and $(\mathrm{P'}_n^\pm)^2=\mathrm{P'}_n^\pm$.
Their key property is that they kill or commute with the fermionic generators depending on the site index:
\begin{alignat}{2}
    &\mathrm{P}_n^+\psi_k^*=\psi_k\mathrm{P}_n^+=0\,, \qquad &k<n\,,\\
    &[\mathrm{P}_n^+,\psi_k^*]=[\mathrm{P}_n^+,\psi_k]=0\,, \qquad &k\geq n\,.
\end{alignat}
\begin{alignat}{2}
    &\mathrm{P}_n^-\psi_k^*=\psi_k\mathrm{P}_n^-=0\,, \qquad &k>n\,,\\
    &[\mathrm{P}_n^-,\psi_k^*]=[\mathrm{P}_n^-,\psi_k]=0\,, \qquad &k\leq n\,.
\end{alignat}
\begin{alignat}{2}
    \label{P'n+}&\psi_k^*\mathrm{P'}_n^+=\mathrm{P'}_n^+\psi_k=0\,, \qquad &k<n\,,\\
    &[\mathrm{P'}_n^+,\psi_k^*]=[\mathrm{P'}_n^+,\psi_k]=0\,, \qquad &k\geq n\,.
\end{alignat}
\begin{alignat}{2}
    &\psi_k^*\mathrm{P'}_n^-=\mathrm{P'}_n^-\psi_k=0\,, \qquad &k>n\,,\\
    &[\mathrm{P'}_n^-,\psi_k^*]=[\mathrm{P'}_n^-,\psi_k]=0\,, \qquad &k\leq n\,.
\end{alignat}
Applying these projectors to the basis states selects precisely the states that correspond to Young diagrams satisfying specific bounds. Concretely, the subspace spanned by the vectors $\mathrm{P}_{m}^-\mathrm{P'}_1^+\ket{\lambda,n+1}$
 The projected space contains only those basis elements for which the Young diagram has length $\ell(\lambda)\leq n$ and first row $\lambda_1\leq m-n$. For an infinite spin chain, the $N$ magnon Hilbert space of the $N$-magnon sector coincides (up to an overall factor) with the space spanned by the states $\mathrm{P'}_1^+\ket{\lambda,N+1}$. The overall factor originates from the ordering ambiguity of the fermionic operators in \eqref{deffreeferfocklambda} and, as we will show, corresponds precisely to the unitary matrix model partition function.
 
Another important algebraic object in the free-fermion construction is the current operator $\mathfrak{J}(z)=\normord\psi^*(z)\psi(z)\normord$ and its Fourier modes:
\begin{align}
    \label{currentoperators}
    \mathfrak{J}_k=\sum_{j\in\mathbb{Z}}\psi_{j+k}^*\psi_j\,,\qquad k\neq0
\end{align}
 for $k=0$, this is simply the total charge operator.

\subsection{The Jordan-Wigner transformation}
As discussed above, the $N$ magnon sector can be obtained from the free-fermion Fock space by a suitable projection. To establish the exact equivalence between the two Hilbert spaces, one also needs a mapping between the operators acting on them that is compatible with the correspondence between the basis vectors. The classic Jordan--Wigner transformation provides precisely such an isomorphism, it reads
\begin{align}
    c_i = (\prod_{j=1}^{i-1} \sigma^z_j)\sigma^-_i\,,\qquad c_i^\dagger = (\prod_{j=1}^{i-1} \sigma^z_j) \sigma^+_i\,,
\end{align}
where $\sigma^{\pm}_i=\frac12(\sigma^x_i\pm i \sigma^y_i)$. The operators $c_i$, $c_i^\dagger$ satisfy the canonical fermionic anticommutation relations
\begin{align}\label{cananticomm}
    \{c_i, c_j^\dagger\}= \delta_{ij}\,.
\end{align}
The factor $\prod_{j=1}^{i-1} \sigma^z_j$ is crucial: it accounts for the ordering of the fermionic operators inherited from the definition of Fock space in free-fermion construction, ensuring that the mapping is an isomorphism between two Hilbert spaces. Physically, this correspondence identifies spin up with a hole (empty site) and spin down with a fermion:
\begin{align}
    \ket{\uparrow}= \ket{0}\,,\qquad \ket{\downarrow}= \ket{1}\,.
\end{align}
Using these fermions, the Hamiltonian of the XXZ spin chain takes the form
\begin{align}
    H_{\text{XXZ}}= \sum_i( c_{i+1}^\dagger c_{i} + c_{i}^\dagger c_{i+1} + \frac{\Delta}{2}(2n_i-1)(2n_{i+1}-1))\,.
\end{align}
where $n_i=c_i^\dagger c_i$ is the number operator at site $i$. For $\Delta=0$ this reduces to a free fermion Hamiltonian. It is clear that the Hamiltonian conserves the total fermion number (equivalently, the number of down spins), consequently, the Hamiltonian is block-diagonal in the magnon sectors introduced earlier, and its eigenstates can be chosen to have a definite number of fermions.  The XYZ chain, on the other hand, does not possess this symmetry and mixes sectors with different fermion numbers.

\section{Classical and quantized symmetric function theory}
\label{Classical and quantized symmetric function theory}

In this section, we give an overview of both classical and quantized symmetric function theories. The theory of symmetric functions has numerous applications in combinatorics, group theory, Lie algebras, and algebraic geometry. The classical theory of symmetric functions deals with symmetric polynomials in commuting variables $\{x_1,\cdots,x_N\}$; the space of such polynomials can be regarded as a linear space equipped with bases labelled by Young tableaux and the associated orthogonality relations. Many different types of bases are known. The classical symmetric functions form an algebraic ring
\begin{align}\label{classicalalgebraicring}
    \mathcal{O}_\mu\mathcal{O}_\nu=\mathcal{C}_{\mu\nu}^\lambda\mathcal{O}_\lambda
\end{align}
where $\mathcal{O}_\mu$ is a basis element labelled by a Young tableau $\mu$ in the variables $\{x_1,\cdots,x_N\}$, and the structure constants $\mathcal{C}_{\mu\nu}^\lambda$ depend only on the choice of basis and on the Young tableaux $\mu$,$\nu$,$\lambda$. 

If we quantize the theory by promoting the commuting variables $\{x_1,\cdots,x_N\}$ to non-commutative operators $\{\hat x_1,\cdots,\hat x_N\}$, and if we can lift each classical basis vector to a corresponding operator, 
\begin{align}
    \mathcal{O}_\mu\to\hat{\mathcal{O}}_\mu\,,
\end{align}
such that the quantized operators obey the product rule
\begin{align}\label{quantizedring}
    \hat{\mathcal{O}}_\mu\hat{\mathcal{O}}_\nu=\hat{\mathcal{C}}_{\mu\nu}^\lambda\hat{\mathcal{O}}_\lambda\,,
\end{align}
with the structure constants unchanged,
\begin{align}\label{quantizedcoefficients}
    \hat{\mathcal{C}}_{\mu\nu}^\lambda=\mathcal{C}_{\mu\nu}^\lambda\,,
\end{align}
for the same basis vectors and Young tableaux, we call the resulting operator algebra a quantized symmetric function theory.

\subsection{Classical symmetric function}
\label{classical symmetric function}
The ring of symmetric polynomials in commuting variables $\{x_1,\cdots,x_N\}$  admits several classical bases, each with its own combinatorial and algebraic significance. The most commonly used are the elementary basis $\{e_\lambda\}$, the complete homogeneous basis $\{h_\lambda\}$, the power-sum basis $\{\powsym_\lambda\}$ and the Schur basis $\{s_\lambda\}$.  A compact way to introduce the first three families is through their generating functions. Let $t$ be a formal indeterminate and define
\begin{align}\label{ehpgeneratingfunction}
    E(t):= \sum_{k\geq 0} t^k\, e_k\,,\qquad H(t):= \sum_{k\geq 0} t^k\, h_k\,,\qquad \Psi(t):=\sum_{k=1}^{\infty} \frac{t^k}{k}  \, \powsym_k \,.
\end{align}
The coefficients \(e_k\) and \(h_k\) are the elementary and complete symmetric functions of degree \(k\); the power-sum symmetric functions are \(\powsym_k = \sum_i x_i^{\,k}\). One also sets \(e_0 = h_0 = 1\). For a partition \(\lambda = (\lambda_1,\dots,\lambda_\ell)\) the corresponding basis elements are defined multiplicatively:
\begin{align}\label{e,h,plambda}
    e_\lambda = e_{\lambda_1}\cdots e_{\lambda_\ell},\qquad
    h_\lambda = h_{\lambda_1}\cdots h_{\lambda_\ell},\qquad
    \powsym_\lambda = \powsym_{\lambda_1}\cdots \powsym_{\lambda_\ell}.
\end{align}
These products determine the ring structure for the \(e\), \(h\) and \(\powsym\) bases, in accordance with the general form \(\mathcal{O}_\mu\mathcal{O}_\nu = \mathcal{C}_{\mu\nu}^\lambda\mathcal{O}_\lambda\) of \eqref{classicalalgebraicring}.

The Schur functions admit an independent combinatorial definition. For a partition \(\lambda\),
\begin{align}
    s_\lambda(x) = \sum_{T \in \mathrm{SSYT}(\lambda)} x^T,
\end{align}
where the sum runs over all semistandard Young tableaux \(T\) of shape \(\lambda\), and \(x^T\) denotes the monomial whose exponent of \(x_i\) counts the occurrences of \(i\) in \(T\). In other words, Schur polynomials are generating functions for SSYTs. Any symmetric polynomial \(f(x)\) can be expanded uniquely in each of these bases:
\begin{align}
    f(x) = \sum_{\lambda} c_\lambda\, \mathfrak{F}_\lambda, \qquad \mathfrak{F}\in\{e,h,\powsym,s\}.
\end{align}
More general families such as Macdonald polynomials, elliptic Macdonald polynomials exist if have another parameters, but we will not require them.

A fundamental relation among the generating functions is
\begin{align}\label{classicaHEP}
    H(t) = e^{\Psi(t)},\qquad E(t) = e^{-\Psi(-t)},
\end{align}
from which the inversion identity
\begin{align}
    E(t)H(-t)=1
\end{align}
immediately follows. These identities will be central when we turn to the quantized version.

The Schur bases also satisfy the celebrated Cauchy formulas:
\begin{align}\label{classicalcauchy}
    \sum_{\lambda} s_\lambda(x) s_\lambda(y) = \prod_{i,j}\frac{1}{1-x_i y_j},\qquad
    \sum_{\lambda} s_\lambda(x) s_{\lambda^T}(y) = \prod_{i,j}(1+x_i y_j).
\end{align}
We will encounter quantum analogs of these equations in the next subsection.

Finally, the algebraic ring structure is encoded in a set of explicit product rules involving Schur functions. For any partitions \(\mu,\nu\) one has
\begin{align}\label{prodshps}
    h_\mu s_\nu &= \sum_{\lambda} K_{\lambda/\nu,\mu}\, s_\lambda,\\
    \powsym_\mu s_\nu &= \sum_{\lambda} \chi^{\lambda/\nu}(\mu)\, s_\lambda,\\
    s_\mu s_\nu &= \sum_{\lambda} C_{\mu\nu}^{\;\;\;\;\lambda}\, s_\lambda,
\end{align}
where \(K_{\lambda/\nu,\mu}\) are the skew Kostka numbers, \(\chi^{\lambda/\nu}(\mu)\) are the skew characters, and \(C_{\mu\nu}^{\lambda}\) are the Littlewood--Richardson coefficients. These relations are the concrete realizations of the abstract structure constants \(\mathcal{C}_{\mu\nu}^\lambda\) introduced in \eqref{classicalalgebraicring}.

\subsection{Quantized symmetric functions}
\label{Quantized symmetric functions}

We now describe a quantization procedure of Fomin and Greene~\cite{fomin1998179} that fulfills the conditions \eqref{quantizedring} and \eqref{quantizedcoefficients}. 

The first step is to promote the commuting variables $\{x_1,\dots,x_N\}$ to non-commuting operators $\{\hat{x}_1,\dots,\hat{x}_N\}$.  The operators \(\hat{x}_i\) are required to satisfy a restricted non-local commutativity:
\begin{align}
   \label{quantizationofvariables} 
   \hat x_i \hat x_j &= \hat x_j \hat x_i, \qquad |i-j|\ge 2,\\ 
   \label{nonlocal2}
   (\hat x_i + \hat x_{i+1})\hat x_i \hat x_{i+1} &= \hat x_i \hat x_{i+1}(\hat x_i + \hat x_{i+1}).
\end{align}
As explained in~\cite{fomin1998179}, these relations encompass several well-known algebras, for instance the plactic algebra, the nilCoxeter algebra and the nil-Temperley--Lieb algebra (the latter can be realized in the free fermion framework discussed in Section~\ref{Quantized symmetric function in spin chain}).

Given a symmetric function \(f(x_1,\dots,x_N)\), there is generally ambiguity in the ordering of the \(\hat{x}_i\) when writing the operator \(\hat{f} = f(\hat{x}_1,\dots,\hat{x}_N)\).  Different choices can still lead to a valid quantized ring, as long as \eqref{quantizedring} and \eqref{quantizedcoefficients} hold.

We adopt the ordering conventions of~\cite{crichigno:2024aub}.  The quantized elementary and complete homogeneous symmetric functions are defined by
\begin{align}\label{ekquantum}
   \hat e_k &:= \sum_{1 \le j_1 < j_2 < \cdots < j_k \le n} \hat x_{j_1} \hat x_{j_2} \cdots \hat x_{j_k}, \qquad 1\le k \le n,\\
\label{hkquantum}
   \hat h_k &:= \sum_{1 \le i_1 \ge i_2 \ge \cdots \ge i_k \le n} \hat x_{i_1} \hat x_{i_2} \cdots \hat x_{i_k}, \qquad 1\le k \le n.
\end{align}
A key consequence of \eqref{quantizationofvariables} is that all quantized elementary symmetric polynomials commute:
\begin{align}\label{commehat}
    \hat e_j \hat e_k = \hat e_k \hat e_j .
\end{align}

The quantized Schur functions are modelled on the classical combinatorial expression.  For a partition \(\lambda\), we set
\begin{align}\label{defShat}
    \hat s_\lambda(\hat x) := \sum_{T} \hat x^T,
\end{align}
where the sum runs over all semistandard Young tableaux \(T\) of shape \(\lambda\), and \(\hat x^T\) denotes the product of the operators \(\hat x_i\) obtained by reading the entries of each column from top to bottom, starting with the leftmost column.\footnote{This column-reading order is the opposite of that used in~\cite{fomin1998179} but agrees with~\cite{crichigno:2024aub}.}
For instance, a semistandard tableau of shape \((2,1)\)
\begin{align*}
   \Yboxdimx{15pt}
   \Yboxdimy{15pt}
   \young({i}{j},{k})
\end{align*}
gives rise to the operator \(\hat x_j \hat x_i \hat x_k\).
With these conventions the one-row and one-column Schur functions coincide with the complete and elementary symmetric functions, respectively:
\begin{align}\label{shse}
    \hat s_{(k)} = \hat h_k, \qquad \hat s_{(1^k)} = \hat e_k .
\end{align}

Because the \(\hat e_k\) commute, the quantized Schur functions inherit the Jacobi--Trudi identity in its determinantal form (Lemma~3.2 of~\cite{fomin1998179}):
\begin{align}\label{Jacobi-Trudi}
    \hat s_{\lambda/\mu} = \det\!\bigl( \hat h_{\lambda_i-\mu_j+j-i} \bigr) = \det\!\bigl( \hat e_{\lambda^T_i-\mu^T_j+j-i} \bigr),
\end{align}
where \(\hat e_0=1\) and \(\hat e_k=0\) for \(k<0\); there is no ordering ambiguity in the determinant because the \(\hat e_k\) commute.  The Jacobi--Trudi formula implies that all quantized Schur functions \(\hat s_\lambda\) commute, so the quantized homogeneous symmetric polynomials commute. We may therefore define the general quantized elementary and homogeneous symmetric polynomials without ordering ambiguity:
\begin{align}\label{quantizedeh}
    \hat{e}_\lambda=\hat{e}_{\lambda_1}\cdots\hat{e}_{\lambda_{\ell(\lambda)}}\,,\qquad \hat{h}_\lambda=\hat{h}_{\lambda_1}\cdots\hat{h}_{\lambda_{\ell(\lambda)}}\,.
\end{align}

A fundamental result of~\cite{fomin1998179} is that whenever the variables obey \eqref{quantizationofvariables}, every algebraic identity among classical Schur functions that can be expressed solely in terms of the \(s_\lambda\) remains valid for the quantized operators \(\hat s_\lambda\).  In particular, the product rules \eqref{quantizedring} hold with the classical structure constants:
\begin{align}\label{hatSSS}
     \hat s_{\mu} \,\hat s_{\nu} &= \sum_{\lambda} C_{\mu\nu}^{\;\;\;\;\lambda} \,\hat s_\lambda,\\
\label{quantizedhs,ss}
     \hat h_{\mu} \,\hat s_{\nu} &= \sum_{\lambda} K_{\lambda/\nu,\mu} \,\hat s_\lambda,
\end{align}
where \(C_{\mu\nu}^{\lambda}\) are the Littlewood--Richardson coefficients and \(K_{\lambda/\nu,\mu}\) the skew Kostka numbers, exactly as in the commutative setting.

The generating functions for the elementary and complete functions are promoted to operators in the obvious way:
\begin{align}
    \hat E(t) := \sum_{k=0}^{\infty} \hat e_k t^k = \prod_{i=1}^{\infty} (1+t\hat x_i), \qquad
    \hat H(t) := \sum_{k=0}^{\infty} \hat h_k t^k = \prod_{i=\infty}^{1} \frac{1}{1-t\hat x_i}.
\end{align}
These quantized generating functions commute with each others and satisfy
\begin{align}\label{hatEH1}
    \hat E(t)\,\hat H(-t) = 1,
\end{align}
and because the \(\hat e_k\) and \(\hat h_k\) commute, all generating functions commute among themselves:
\[
    [\hat E(t),\hat E(t')] = [\hat H(t),\hat H(t')] = [\hat E(t),\hat H(t')] = 0 .
\]

The quantized power-sum symmetric functions \(\hat \powsym_k\) are not expressed as a simple polynomial in the \(\hat x_i\), instead they are defined implicitly through the logarithm of the complete generating function, similar as the classical relation \eqref{classicaHEP}:
\begin{align}
    \hat \Psi(t) := \sum_{k=1}^{\infty} \frac{\hat \powsym_k}{k}\, t^k = \log \hat H(t).
\end{align}
This definition guarantees that the operator analog of \eqref{classicaHEP} holds, \(\hat H(t) = \exp(\hat\Psi(t))\), and that the ring structure for products involving \(\hat \powsym_\mu\) will reproduce the classical skew character expansion when acting on Schur functions.

Finally, the classical Cauchy identities \eqref{classicalcauchy} admit a non-commutative counterpart~\cite{fomin1998179}.  Let \(t_i\) denote ordinary commuting variables and \(\hat y_j\) operators satisfying \eqref{quantizationofvariables}.  Then
\begin{align}
    \sum_{\lambda} s_\lambda(t) \,\hat s_{\lambda}(\hat y) = \prod_{i=1}^m \hat H(t_i), \label{NC-Cauchy}
\end{align}
where the sum runs over all partitions \(\lambda\).  This identity will be instrumental in the construction of off-shell Bethe states and Toda chain tau functions later in the paper.

\subsection{Quantized symmetric functions in the spin chain}
\label{Quantized symmetric function in spin chain}

We now show that the free fermion algebra provides a concrete realization of the quantized variables satisfying \eqref{quantizationofvariables} and allows us to construct the quantized power-sum symmetric functions explicitly.  Consider a semi-infinite set of fermionic modes \(c_i, c_i^\dagger\) obeying the canonical anti-commutation relations \eqref{cananticomm} and the projection onto the Hilbert space described in Section~\ref{Heisenberg spin chains and Free fermions}.  Define the elementary shift operators
\begin{align}\label{defxi}
    \hat x_i = c_{i+1}^\dagger c_i\,, \qquad \hat x_i^\dagger = c_i^\dagger c_{i+1}\,, \qquad i\ge 1\,.
\end{align}
These operators move a fermion from site \(i\) to site \(i+1\) or change $\ket{\downarrow}_i$ to $\ket{\uparrow}_i$ and $\ket{\uparrow}_{i+1}$ to $\ket{\downarrow}_{i+1}$ and vice-versa; they act on a semi-infinite chain, so they are infinite-dimensional.  A direct calculation shows that they satisfy the commutation relations required by the Fomin Greene quantization scheme, namely \eqref{quantizationofvariables}.  Moreover, because a fermion cannot be created twice on the same site, one has the nilpotency condition
\begin{align}\label{nilpotx}
    \hat x_i^2 = (\hat x_i^\dagger)^2 = 0\, .
\end{align}
With this additional constraint the algebra of the \(\hat x_i\) reduces to the nil-Temperley--Lieb algebra mentioned in Section~\ref{Quantized symmetric functions}.  Consequently, the relations \eqref{nonlocal2} simplify to
\begin{align}\label{algebraicinspinchain}
    \hat x_i \hat x_j &= \hat x_j \hat x_i\,, \qquad |i-j|\ge 2, \\
    \hat x_{i+1}\hat x_i\hat x_{i+1} &= \hat x_i \hat x_{i+1}\hat x_i \,.
\end{align}
These are precisely the braid relations satisfied by the elementary transpositions \(s_i\) of the symmetric group \(S_N\), with the only difference that \(s_i^2=1\) while here \(\hat x_i^2=0\); this is nothing but nil-Temperley--Lieb algebra.

A central role is played by the current operators, which are obtained from the general free fermion currents \(\mathfrak{J}_k\) by the projection \(\mathrm{P'}^+_1\) that isolates the semi-infinite chain:
\begin{align}
    \hat J_k = \sum_{i\ge 1} c_{i+k}^\dagger c_i = \mathrm{P'}^+_1 \mathfrak{J}_k\,, \qquad
    \hat J_k^\dagger = \sum_{i\ge 1} c_i^\dagger c_{i+k} = \mathfrak{J}_{-k} \mathrm{P'}^+_1, \qquad k\in \mathbb{Z}_{\geq 0}\, .
\end{align}
The operator \(\hat J_k\) shifts a fermion \(k\) steps to the right, picking up the appropriate fermionic signs when hopping over occupied sites, while \(\hat J_k^\dagger\) shifts it to the left.  On the \(N\) magnon vacuum they act as
\begin{align}
    \hat J_k^\dagger \ket{\emptyset,N} = 0\,, \qquad \bra{\emptyset,N} \hat J_k = 0\, .
\end{align}
These currents satisfy the commutation relations
\begin{align}
    [\hat J_k, \hat J_{k'}] = 0, \qquad [\hat J^\dagger_k, \hat J^\dagger_{k'}] = 0\,,
\end{align}
while the mixed commutator takes the simple form
\begin{align}
    [\hat J^\dagger_{k+m}, \hat J_k] = \sum_{i=1}^k c^\dagger_i c_{i+m}\,, \qquad
    [\hat J^\dagger_k, \hat J_{k+m}] = \sum_{i=1}^k c^\dagger_{i+m} c_i\,, \qquad m\ge 0\,.
\end{align}

The generating function of the quantized complete symmetric functions can be written as an exponential of the currents.  Using the Baker--Campbell--Hausdorff formula one finds
\begin{align}\label{Hvertex}
    \hat H(t) = \exp\!\Bigl( \sum_{k=1}^{\infty} \frac{t^k}{k} \hat J_k \Bigr)\,.
\end{align}
Comparing with the classical relation \eqref{classicaHEP} suggests that the currents are the natural quantized power-sum symmetric functions, we therefore set
\begin{align}
    \hat{\powsym}_k = \hat J_k\,, \qquad \hat{\powsym}_k^\dagger = \hat J_k^\dagger\, .
\end{align}
With this identification the algebraic structure required by \eqref{quantizedring} is fully realized for the \((e,h,\powsym,s)\) family.

We now verify the algebraic ring relations.  The quantized elementary and complete symmetric functions are defined in~\eqref{quantizedeh}, and the quantized Schur functions by the combinatorial formula \eqref{defShat}.  Because the operators \(\hat x_i\) obey \eqref{quantizationofvariables} and \eqref{nonlocal2}, all identities involving only quantized elementary, homogeneous and Schur symmetric functions, and expressible solely in terms of the \(\hat s_\lambda\), hold exactly as in the commutative theory by \eqref{shse}.  In particular, one has the product formulas
\begin{align}\label{prodshpsquantum}
    \hat h_\mu \hat s_\nu &= \sum_{\lambda} K_{\lambda/\nu,\mu}\, \hat s_\lambda, \\
    \hat s_\mu \hat s_\nu &= \sum_{\lambda} C_{\mu\nu}^{\;\;\;\;\lambda}\, \hat s_\lambda .
\end{align}
For the quantized power-sum polynomials the analogous relation reads
\begin{align}\label{quantizedps}
    \hat{\powsym}_\mu \hat s_\nu = \sum_{\lambda} \chi^{\lambda/\nu}(\mu)\, \hat s_\lambda,
\end{align}
where \(\chi^{\lambda/\nu}(\mu)\) denotes the skew character of the symmetric group.  Combining \eqref{quantizedps} with \eqref{shse} and the definitions of the quantized elementary and homogeneous functions, one sees that the full set of structure constants coincides with its classical counterpart, proving that the quantized ring satisfies \eqref{quantizedcoefficients} for the bases \((e,h,\powsym,s)\).

A key geometric interpretation arises from the action of these operators on the \(N\) magnon vacuum.  The state corresponding to the empty Young diagram is
\begin{align}
    \ket{\emptyset,N} = \ket{\downarrow^N \uparrow \cdots},
\end{align}
and from the definition of \(\hat s_\lambda\) together with \eqref{defxi} it follows that
\begin{align}\label{svacyoung}
    \hat s_\lambda \ket{\emptyset,N} = \ket{\lambda,N},
\end{align}
where \(\ket{\lambda,N}\) is the basis vector associated with the Young diagram \(\lambda\) in the \(N\) magnon sector, and the proof is attached in~\ref{Proof ofeqrefsvacyoung}.  The state vanishes identically if the length \(\ell(\lambda)\) exceeds \(N\), showing that the choice of the vacuum automatically restricts the admissible Young diagrams.  Acting with the product identities \eqref{prodshpsquantum} and \eqref{quantizedps} on a generic basis vector then yields the explicit action
\begin{align}\label{quantizedhpsactonnu}
    \hat h_\mu \ket{\nu} &= \sum_{\lambda} K_{\lambda/\nu,\mu} \ket{\lambda}, \\
    \hat{\powsym}_\mu \ket{\nu} &= \sum_{\lambda} \chi^{\lambda/\nu}(\mu) \ket{\lambda},\label{quantizedpmunu} \\
    \hat s_\mu \ket{\nu} &= \sum_{\lambda} C_{\mu\nu}^{\;\;\;\;\lambda} \ket{\lambda}.
\end{align}
Thus the quantized symmetric functions encode the skew Kostka numbers, the skew characters, and the Littlewood--Richardson coefficients directly in the spin-chain Hilbert space. The proof of~\eqref{quantizedps} and~\eqref{quantizedpmunu} is given in appendix~\ref{Proof ofeqrefquantizedps}.

We note that for a finite chain of length \(n\) one simply sets
\begin{align}\label{xiopen}
    \hat x_i = 0 \qquad \text{for all } i \ge n,
\end{align}
which truncates the algebra consistently.  In the unitary matrix model applications discussed in the next section, however, we always consider semi-infinite chain.

The fact that all structure constants are preserved means that the ring of quantized symmetric functions provides a faithful operator realization of the classical theory.  This observation is the central ingredient that will allow us, in the following, to map a large class of unitary matrix integrals to vacuum correlation functions of these operators in the spin-chain Hilbert space.

\section{Unitary matrix models and quantized symmetric functions}
\label{Unitary matrix model and quantized symmetric function}

We now apply the formalism of quantized symmetric functions to unitary matrix integrals.  A unitary matrix model is a random matrix theory defined by integration over the unitary group \(U(N)\).  Its partition function has the general form
\[
    Z_N = \int_{U(N)} \mathrm{d}M \; \exp\bigl(-\operatorname{Tr} V(M)\bigr),
\]
and such models play a central role in two dimensional quantum gravity, gauge theories and string theory.  More generally, one considers integrals of gauge invariant functions
\[
    \int \mathrm{d}M \, f(M).
\]

After diagonalisation \(M = A \Lambda A^\dagger\) with \(\Lambda = \operatorname{diag}(\lambda_1,\dots,\lambda_N)\) and \(A\in U(N)\), the integral reduces to an integral over the eigenvalues on the unit circle:
\begin{align}\label{unimatrixmodelintegral}
    \int \mathrm{d}M \, f(M) = \frac{1}{N!} \int_{\mathbb{T}^N} \prod_{i=1}^N \frac{\mathrm{d}\lambda_i}{2\pi\mathrm{i}\,\lambda_i}\; |\Delta(\lambda)|^2 \, f(\lambda),
\end{align}
where \(\Delta(\lambda) = \prod_{i<j} (\lambda_i - \lambda_j)\) is the Vandermonde determinant.  The factor \(|\Delta(\lambda)|^2\) is the Jacobian of the diagonalisation, and the prefactor \(1/N!\) accounts for the residual Weyl symmetry.  Because the function \(f(M)\) is gauge invariant, its eigenvalue version \(f(\lambda)\) is symmetric under any permutation of the \(\lambda_i\). In order to translate the integral into the language of symmetric functions, we need to assume that the integrand can be written as a sum of products of symmetric polynomials in the $\{\lambda_i\}$ and symmetric polynomials in their inverses $\{\lambda_i^{-1}\}$.
\begin{align}\label{assoff}
    f(\lambda) = \sum_{a=1}^\infty \mathfrak{p}_a(\lambda_1,\dots,\lambda_N)\; \mathfrak{p}'_a(\lambda_1^{-1},\dots,\lambda_N^{-1}),
\end{align}
where \(\mathfrak{p}_a\) and \(\mathfrak{p}'_a\) are symmetric polynomials, expanding each term in the Schur basis,
\[
    \mathfrak{p}_a(\lambda) = \sum_{\ell(\mu)\le N} c^+_{a,\mu}\, s_\mu(\lambda), \qquad
    \mathfrak{p}'_a(\lambda^{-1}) = \sum_{\ell(\nu)\le N} c^-_{a,\nu}\, s_\nu(\lambda^{-1}),
\]
and using the orthogonality of the Schur functions on the unitary group,
\begin{align}
    \frac{1}{N!} \int_{\mathbb{T}^N} \prod_{i=1}^N \frac{\mathrm{d}\lambda_i}{2\pi\mathrm{i}\,\lambda_i}\; |\Delta(\lambda)|^2 \, s_\mu(\lambda) \, s_\nu(\lambda^{-1}) = \delta_{\mu\nu},
\end{align}
the integral collapses to a sum of products of the expansion coefficients:
\begin{align}
    \int \mathrm{d}M \, f(M) = \sum_{a=1}^\infty \sum_{\ell(\lambda)\le N} c^+_{a,\lambda}\, c^-_{a,\lambda}.
\end{align}

The key observation is that this algebraic structure can be lifted to the quantized setting.  According to the correspondence between classical symmetric functions and their quantized counterparts established in the previous sections, we replace every Schur function \(s_\lambda\) by the operator \(\hat s_\lambda\) acting on the spin-chain Hilbert space, and the integral with the squared Vandermonde determinant as the measure can be reinterpreted as the vacuum expectation value in the \(N\) magnon sector.  More precisely, if \(f(\lambda)\) admits the factorized form \eqref{assoff}, the matrix integral is mapped to the correlation function
\begin{align}\label{matrixmodelspinchaincorrespondence}
    \frac{1}{N!} \int_{\mathbb{T}^N} \prod_{i=1}^N \frac{\mathrm{d}\lambda_i}{2\pi\mathrm{i}\,\lambda_i}\; |\Delta(\lambda)|^2 \, f(\lambda)
    \;\longleftrightarrow\;
    \bra{\emptyset,N} \sum_{a=1}^\infty \hat{\mathfrak{p}}'^\dagger_a \, \hat{\mathfrak{p}}_a \ket{\emptyset,N}
    = \bra{\emptyset,N} \, \normord\hat f\normord \, \ket{\emptyset,N},
\end{align}
where \(\normord\cdots\normord\) denotes normal ordering (all adjoint operators moved to the left).  The operator \(\hat f\) is obtained from \(f(\lambda)\) by lifting the classical symmetric polynomials $\mathfrak{p}_a(\lambda)$, $\mathfrak{p}'_a(\lambda)$ to quantized version $\hat{\mathfrak{p}}_a$, $\hat{\mathfrak{p}}'^\dagger_a$. For instance 
\begin{align}\label{trm^ktopk}
    \operatorname{Tr}(M^k) \;\longrightarrow\; \hat{\powsym}_k, \qquad
    \operatorname{Tr}(M^{\dagger k}) \;\longrightarrow\; \hat{\powsym}_k^\dagger .
\end{align}
This mapping is the central result of the paper: it translates a wide class of unitary matrix models into vacuum correlators of quantized symmetric functions.

\subsection{GWW model}\label{GWW model}
A simple and instructive example is provided by the Gross--Witten--Wadia
(GWW) model, whose partition function is
\[
    Z_N = \int_{U(N)} \mathcal{D}M \; \exp\!\Bigl( -\frac{N}{\lambda} \bigl( \operatorname{Tr} M + \operatorname{Tr} M^\dagger \bigr) \Bigr)\,.
\]
Under the quantization map \eqref{trm^ktopk} this becomes
\[
    Z_N = \bra{\emptyset,N} \normord\exp\Bigl( -\frac{N}{\lambda} \bigl( \hat p_1 + \hat p_1^\dagger \bigr) \Bigr) \normord\ket{\emptyset,N}=\bra{\emptyset,N} \mathrm{e}^C\ket{\emptyset,N}\,.
\]
where 
\begin{align}
    C=-\frac{N}{\lambda}(\hat \powsym_1+\hat \powsym^\dagger_1)+\frac{N^2}{2\lambda^2}n_1+\frac{N^3}{12\lambda^3}(c_1^\dagger c_2+c_2^\dagger c_1)+\frac{N^4}{24\lambda^4}(n_2-n_1)+\mathcal{O}\left((\frac{N}{\lambda})^5\right)\,.
\end{align}
The higher order terms in \(C\) encode the corrections induced by normal
ordering. Physically, the parameter \(N/\lambda\) plays the role of an
Euclidean evolution time. Thus the small \(N/\lambda\) limit probes the
short time dynamics of the spin chain around the \(N\)-magnon vacuum, where
only nearest neighbour hopping contributes at leading order. In this regime
the matrix integral is controlled by the free-fermion XX dynamics, while the
normal ordering corrections generate local density and hopping deformations at
higher orders in \(N/\lambda\).

This gives a concrete spin chain interpretation of the weak coupling expansion
of the GWW model. It is also consistent with the broader idea that matrix model
phase structure can be translated into spin chain dynamics. Related connections
between AdS black hole entropy and spin chain dynamics have been discussed
in~\cite{Perez-Garcia:2024pcq}.

\subsection{superconformal indices of $\mathcal{N}=4$ SYMs}\label{superconformal indices of N=4 SYMs}
Superconformal index is a key quantitative tool for probing AdS/CFT correspondence. To rewrite superconformal indices of $\mathcal{N}=4$ SYMs as an correlation function between spin chain vacuum, we begin with the matrix integral representation of the $\mathcal{N}=4$ superconformal index for the $U(N)$ gauge group of SYM theory. As established in~\cite{Kinney_2007}, after adjusting the fugacities to match our notation, the index takes the form
\begin{equation}
\begin{aligned}
I_N(t, u, v;p, q)=\int_{U(N)} dU  \exp \left( \sum_{n=1}^\infty \frac{f(t^n, u^n, v^n; p^n, q^n)}{n} \mathrm{Tr} (U^n) \mathrm{Tr} [ (U^\dagger)^n ] \right),
\label{eq:SCI}
\end{aligned}
\end{equation}
where $f(t, u, v;p, q)$ denotes the single-letter index of the theory, expressed as
\begin{align}\label{eq:single-letter-index}
f(t, u, v; p, q)=1-\frac{(1-t)(1-u)(1-v)}{(1-p)(1-q)}.
\end{align}
After diagonalization, we have:
\begin{equation}
\begin{aligned}
I_N(t, u, v; p, q)&=\frac{1}{N!} \oint_{\mathbb{T}^N} \prod_{i=1}^N \frac{dx_i}{2\pi \ii x_i} \prod_{1 \leq i \ne j \leq N} \left( 1-\frac{x_i}{x_j} \right)\\
&\quad \times \exp \left( \sum_{n=1}^\infty \frac{f(t^n, u^n, v^n; p^n, q^n)}{n} \powsym_n(\x) \powsym_n(\x^{-1}) \right),
\end{aligned}
\label{eq:SCI-2}
\end{equation}
introducing the shorthand \(f_n = f(t^n, u^n, v^n; p^n, q^n)\), then the integrand is expanded as
\begin{align}
    \e\left(\sum_{n=1}^\infty \frac{f_n}{n} \powsym_n(\x) \powsym_n(\x^{-1})\right)=\sum_{\mu}\frac{f_\mu}{z_\mu}\powsym_\mu(\x)\powsym_\mu(\x^{-1})\,.
\end{align}
The power-sum symmetric functions are expanded in the Schur basis by the Frobenius formula,
\begin{align}
    \powsym_\mu(\x)=\sum_{\substack{\lambda\vdash n\\\ell(\lambda)\leq N}}\chi_{\mu}^{\lambda}s_\lambda(\x)\,.
\end{align}
Using the orthonormality of the Schur functions on the unitary group,
\begin{align}
    I_N(t, u, v; p, q)=\sum_{\mu}\frac{f_\mu}{z_\mu}\sum_{\substack{\lambda\vdash n\\\ell(\lambda)\leq N}}(\chi_{\mu}^{\lambda})^2\,.
\end{align}
According to the general dictionary~\eqref{matrixmodelspinchaincorrespondence}, this result translates directly into a vacuum expectation value of the quantized power-sum operators:
\begin{align}\label{SCICorrelationfunction}
   I_N(t, u, v; p, q)= \bra{\emptyset,N}\normord\e\left(\sum_{n=1}^{\infty}\frac{f_n}{n}\hat{\powsym}^\dagger_n\hat{\powsym}_n\right)\normord\ket{\emptyset,N}
\end{align}
where the normal ordering moves all \(\hat{\powsym}_n^\dagger\) to the left of all \(\hat{\powsym}_n\) in any product.

The same formalism can be applied to the superconformal indices of \(\mathcal{N}=4\) SYM with \(Sp(2N)\), \(SO(2N+1)\) and \(SO(2N)\) gauge groups, 
by exploiting the characteristic trace relations and the unitarity of these groups. To do this, we first introduce the integral representation of these superconformal indices of $\mathcal{N}=4$ SYMs~\cite{Spiridonov:2010qv}:
\begin{align}
\label{SO(2N)SCI}
&I_{SO(2N)}^{SCI} = \chi'_N \oint_{\mathbb{T}^N} 
\prod_{1 \leq i < j \leq N} \frac{\Gamma(t z_i^{\pm 1} z_j^{\pm 1}; p, q)\Gamma(u z_i^{\pm 1} z_j^{\pm 1}; p, q)}{\Gamma(z_i^{\pm 1} z_j^{\pm 1}; p, q)\Gamma(tu z_i^{\pm 1} z_j^{\pm 1}; p, q)} 
\prod_{j=1}^N \frac{dz_j}{2\pi \ii z_j},\\
\label{Sp(2N)SCI}
&I_{Sp(2N)}^{SCI} = \chi'_N \oint_{\mathbb{T}^N} 
\prod_{1 \leq i < j \leq N} \frac{\Gamma(t z_i^{\pm 1} z_j^{\pm 1},u z_i^{\pm 1} z_j^{\pm 1}; p, q)}{\Gamma(z_i^{\pm 1} z_j^{\pm 1},tu z_i^{\pm 1} z_j^{\pm 1}; p, q)} 
\prod_{j=1}^N \frac{\Gamma(t z_j^{\pm 2},u z_j^{\pm 2}; p, q)}{\Gamma(z_j^{\pm 2},tu z_j^{\pm 2}; p, q)} 
\frac{dz_j}{2\pi \ii z_j},\\
\label{SO(2N+1)SCI}
&I_{SO(2N+1)}^{SCI} = \chi'_N \oint_{\mathbb{T}^N} 
\prod_{1 \leq i < j \leq N} \frac{\Gamma(t z_i^{\pm 1} z_j^{\pm 1},u z_i^{\pm 1} z_j^{\pm 1}; p, q)}{\Gamma(z_i^{\pm 1} z_j^{\pm 1},tu z_i^{\pm 1} z_j^{\pm 1}; p, q)} 
\prod_{j=1}^N \frac{\Gamma(t z_j^{\pm 1},u z_j^{\pm 1}; p, q)}{\Gamma(z_j^{\pm 1},tu z_j^{\pm 1}; p, q)} 
\frac{dz_j}{2\pi \ii z_j},
\end{align}
where
\begin{equation*}
\chi'_N  = \left\{
    \begin{aligned}
    & \frac{(p;p)_\infty^N (q;q)_\infty^N\Gamma^N(t; p, q)\Gamma^N(u; p, q)}{2^{N}N!\Gamma^N(tu; p, q)} \ && \text{$Sp(2N)$ or $SO(2N+1)$}\,, \\
    & \frac{(p; p)_\infty^N (q; q)_\infty^N\Gamma^N(t; p, q)\Gamma^N(u; p, q)}{2^{N-1} N!\Gamma^N(tu; p, q)}  && \text{$SO(2N)$}\,. \\
    \end{aligned}
\right.
\end{equation*}
The definition of elliptic gamma function $\Gamma(x;p,q)$ is
\begin{align}\label{ellgammadef}
    \Gamma(x;p,q)=\e\bigg[\sum_{n=1}^\infty\frac{x^n-(pq)^nx^{-n}}{n(1-p^n)(1-q^n)}\bigg]\,,
\end{align}
and
\begin{align}
\begin{split}
       &\Gamma(a,b;p,q)=\Gamma(a;p,q)\Gamma(b;p,q)\,,\qquad
       \Gamma(a^{\pm n};p,q)=\Gamma(a^{+ n};p,q)\Gamma(a^{- n};p,q)\,,\\
       &\Gamma(a^{\pm n}b^{\pm m};p,q)=\Gamma(a^{+ n}b^{+ m};p,q)\Gamma(a^{+ n}b^{- m};p,q)\Gamma(a^{- n}b^{+ m};p,q)\Gamma(a^{- n}b^{- m};p,q)\,.
\end{split}
\end{align}
The integrands can be rewritten in an exponential form similar to the \(U(N)\) case. 
Introducing the auxiliary expressions
then the integrand can be rewritten as
\begin{align}\label{BCD_nauxiliary}
\begin{split}
        C_n&=\sum_{1\leq i<j\leq N}\bigg[(z_i/z_j)^n+(z_j/z_i)^n+(z_iz_j)^n+(z_iz_j)^{-n}\bigg]+\sum_{i=1}^N \bigg[z_i^{2n}+z_i^{-2n}\bigg]+N\,,\\
        B_n&=\sum_{1\leq i<j\leq N}\bigg[(z_i/z_j)^n+(z_j/z_i)^n+(z_iz_j)^n+(z_iz_j)^{-n}\bigg]+\sum_{i=1}^N \bigg[z_i^{n}+z_i^{-n}\bigg]+N\,,\\
        D_n&=\sum_{1\leq i<j\leq N}\bigg[(z_i/z_j)^n+(z_j/z_i)^n+(z_iz_j)^n+(z_iz_j)^{-n}\bigg]+N\,,
\end{split}
\end{align}
we have
\begin{align}
    &I_{Sp(2N)}^{SCI}=\frac{1}{2^N N!}\oint_{\mathbb{T}^N}\prod_{j=1}^N\frac{dz_j}{2\pi \ii z_j}\prod_{1 \leq i \ne j \leq N} \left( 1-\frac{z_i}{z_j} \right)\times\exp\bigg[\frac{\powsym_n(\z)\powsym_n(\z^{-1})}{n}\bigg]\notag\\
    &\quad\quad\quad\quad\quad\quad\quad\quad\quad\times \exp\bigg[-\sum_{n=1}^\infty\frac{(1-t^n)(1-u^n)(1-(pq/tu)^n)C_n(\z,\z^{-1})}{n(1-p^n)(1-q^n)}\bigg]\,,\\
    &I_{SO(2N+1)}^{SCI}=\frac{1}{2^N N!}\oint_{\mathbb{T}^N}\prod_{j=1}^N\frac{dz_j}{2\pi \ii z_j}\prod_{1 \leq i \ne j \leq N} \left( 1-\frac{z_i}{z_j} \right)\times\exp\bigg[\frac{\powsym_n(\z)\powsym_n(\z^{-1})}{n}\bigg]\notag\\
    &\quad\quad\quad\quad\quad\quad\quad\quad\quad\times \exp\bigg[-\sum_{n=1}^\infty\frac{(1-t^n)(1-u^n)(1-(pq/tu)^n)B_n(\z,\z^{-1})}{n(1-p^n)(1-q^n)}\bigg]\,,\\
    &I_{SO(2N)}^{SCI}=\frac{1}{2^{N-1} N!}\oint_{\mathbb{T}^N}\prod_{j=1}^N\frac{dz_j}{2\pi \ii z_j}\prod_{1 \leq i \ne j \leq N} \left( 1-\frac{z_i}{z_j} \right)\times\exp\bigg[\frac{\powsym_n(\z)\powsym_n(\z^{-1})}{n}\bigg]\notag\\
    &\quad\quad\quad\quad\quad\quad\quad\quad\quad\times \exp\bigg[-\sum_{n=1}^\infty\frac{(1-t^n)(1-u^n)(1-(pq/tu)^n)D_n(\z,\z^{-1})}{n(1-p^n)(1-q^n)}\bigg]\,,
\end{align}
using the fact that
\begin{align}
    \bigg(\powsym_n(\z)+\powsym_n(\z^{-1})\bigg)\bigg(\powsym_n(\z)+\powsym_n(\z^{-1})\bigg)&=2\sum_{1\leq i<j\leq N}\bigg[(z_i/z_j)^n+(z_j/z_i)^n+(z_iz_j)^n+(z_iz_j)^{-n}\bigg]\notag\\
    &+\sum_{i=1}^N \bigg[z_i^{2n}+z_i^{-2n}\bigg]+2N\,,
\end{align}
it is easy to prove
\begin{align}
\begin{split}
    \sum_{1\leq i<j\leq N}&\bigg[(z_i/z_j)^n+(z_j/z_i)^n+(z_iz_j)^n+(z_iz_j)^{-n}\bigg]+N=\\
    &\frac{1}{2}\bigg(\powsym_n(\z)+\powsym_n(\z^{-1})\bigg)\bigg(\powsym_n(\z)+\powsym_n(\z^{-1})\bigg)-\frac{1}{2}\bigg(\powsym_{2n}(\z)+\powsym_{2n}(\z^{-1})\bigg)\,,
\end{split}
\end{align}
according to this, we can written the~\eqref{BCD_nauxiliary} in power-sum symmetric functions
\begin{align}
    \begin{split}
         C_n&=\frac{1}{2}\bigg(\powsym_n(\z)+\powsym_n(\z^{-1})\bigg)\bigg(\powsym_n(\z)+\powsym_n(\z^{-1})\bigg)+\frac{1}{2}\bigg(\powsym_{2n}(\z)+\powsym_{2n}(\z^{-1})\bigg)\,,\\
        B_n&=\frac{1}{2}\bigg(\powsym_n(\z)+\powsym_n(\z^{-1})+2\bigg)\bigg(\powsym_n(\z)+\powsym_n(\z^{-1})\bigg)-\frac{1}{2}\bigg(\powsym_{2n}(\z)+\powsym_{2n}(\z^{-1})\bigg)\,,\\
        D_n&=\frac{1}{2}\bigg(\powsym_n(\z)+\powsym_n(\z^{-1})\bigg)\bigg(\powsym_n(\z)+\powsym_n(\z^{-1})\bigg)-\frac{1}{2}\bigg(\powsym_{2n}(\z)+\powsym_{2n}(\z^{-1})\bigg)\,.
    \end{split}
\end{align}
To find the corresponding correlation functions in spin chain, we should lift~\eqref{BCD_nauxiliary} to quantum operators, so we define
\begin{align}
    \begin{split}
        \hat{A}_n&=\hat{\powsym}_n\hat{\powsym}_n^\dagger\,,\\
        \hat{C}_n&=\frac{1}{2}(\hat{\powsym}_n+\hat{\powsym}_n^\dagger)^2+\frac{1}{2}(\hat{\powsym}_{2n}+\hat{\powsym}_{2n}^\dagger)\,,\\
        \hat{B}_n&=\frac{1}{2}(\hat{\powsym}_n+\hat{\powsym}_n^\dagger)^2-\frac{1}{2}(\hat{\powsym}_{2n}+\hat{\powsym}_{2n}^\dagger-2\hat{\powsym}_{n}-2\hat{\powsym}_{n}^\dagger)\,,\\
        \hat{D}_n&=\frac{1}{2}(\hat{\powsym}_n+\hat{\powsym}_n^\dagger)^2-\frac{1}{2}(\hat{\powsym}_{2n}+\hat{\powsym}_{2n}^\dagger)\,.
    \end{split}
\end{align}
Substituting these into the partition functions and recalling the dictionary~\eqref{matrixmodelspinchaincorrespondence}, we finally obtain the superconformal indices as spin-chain vacuum correlators:
\begin{align}
    \begin{split}
       I_{U(N)}^{SCI}&=\bra{\emptyset,N}\normord\e\bigg[\sum_{n=1}^{\infty}\bigg(\frac{\hat{A}_n}{n}-\frac{g_n}{n}\hat{A}_n\bigg)\bigg]\normord\ket{\emptyset,N}\,,\\
       I_{Sp(2N)}^{SCI}&=\frac{1}{2^N}\bra{\emptyset,N}\normord\e\bigg[\sum_{n=1}^{\infty}\bigg(\frac{\hat{A}_n}{n}-\frac{g_n}{n}\hat{C}_n\bigg)\bigg]\normord\ket{\emptyset,N}\,,\\
    I_{SO(2N)}^{SCI}&=\frac{1}{2^{N-1}}\bra{\emptyset,N}\normord\e\bigg[\sum_{n=1}^{\infty}\bigg(\frac{\hat{A}_n}{n}-\frac{g_n}{n}\hat{D}_n\bigg)\bigg]\normord\ket{\emptyset,N}\,,\\
    I_{SO(2N+1)}^{SCI}&=\frac{1}{2^N}\bra{\emptyset,N}\normord\e\bigg[\sum_{n=1}^{\infty}\bigg(\frac{\hat{A}_n}{n}-\frac{g_n}{n}\hat{B}_n\bigg)\bigg]\normord\ket{\emptyset,N}\,,
    \end{split}
\end{align}
where
\begin{align}
    g_n=\frac{(1-t^n)(1-u^n)(1-(pq/tu)^n)}{(1-p^n)(1-q^n)}\,.
\end{align}

\subsection{Toda tau functions}

The unitary matrix integral that appears in Ref.~\cite{murthy:2022ien},
\begin{align}\label{Todamatrixmodel}
    \tilde{Z}_N(\mathbf{t}^+,\mathbf{t}^-) = \int_{U(N)} \mathrm{d}U \,
    \exp\!\Bigg( \sum_{k=1}^\infty \frac{1}{k} \bigl( t_k^+ \operatorname{Tr} U^k + t_k^- \operatorname{Tr} U^{-k} \bigr) \Bigg),
\end{align}
can be regarded as a Hirota tau function of Toda type.  General Hirota tau functions admit a free-fermion construction~\cite{alexandrov:2012tr} (or an equivalent phase model construction using the isomorphism map introduced in~\cite{araujo:2024klz}) in terms of group-like elements.  In this language, Toda tau functions are expressed as inner products of off-shell Bethe states in the phase model.

\subsubsection{Bosonic chain}

Consider a bosonic chain of length \(M+1\).  Its Hilbert space is spanned by the Fock states
\begin{align}
    \ket{n_1,n_2,\dots,n_{M+1}} = \ket{n_1}_1 \otimes \ket{n_2}_2 \otimes \cdots \otimes \ket{n_{M+1}}_{M+1},
\end{align}
where \(n_j \ge 0\) denotes the number of bosons at site \(j\).

The \(q\)-boson model is a well-known bosonic chain whose Hamiltonian reads
\begin{align}
    H_B = -\sum_{j=1}^{M+1} \bigl( \phi_j^\dagger \phi_{j+1} + \phi_{j+1}^\dagger \phi_j - 2 N_j \bigr),
\end{align}
where \(N_j\) is the local boson number operator, and the operators \(\phi_i,\phi_i^\dagger\) satisfy the q-deformed Heisenberg commutation relation and act on the site basis as
\begin{align}
   \phi_i \ket{n}_i =
   \begin{cases}
      \ket{n-1}_i, & \text{for } n>0,\\[2pt]
      0,           & \text{for } n=0,
   \end{cases}
   \qquad
   \phi_i^\dagger \ket{n}_i = \ket{n+1}_i .
\end{align}
In the limit \(q\to\infty\) the model reduces to the \emph{phase model}, which has the same Hamiltonian; the parameter \(q\) then only appears in the commutation relations of the bosonic operators.

In the phase model of length \(M+1\) we consider the \(N\) particles off-shell Bethe state \(\ket{y_1,\dots,y_N}\) constructed within the algebraic Bethe ansatz (``off-shell'' means that the Bethe coordinates \(y_i\) are not required to satisfy the Bethe equations)~\cite{araujo:2024klz}.  The inner product of two such states defines the function
\begin{align}
    \mathcal{I}(N,M\mid \mathbf{x},\mathbf{y}) = \braket{\mathbf{x}}{\mathbf{y}} .
\end{align}
Introduce the Miwa coordinates \(\mathbf{t}^+ = (t_1^+,t_2^+,\dots)\) and \(\mathbf{t}^- = (t_1^-,t_2^-,\dots)\) via
\begin{align}
    t_k^+ = \frac{1}{k} \sum_{i=1}^N x_i^{\,k}, \qquad
    t_k^- = \frac{1}{k} \sum_{i=1}^N y_i^{\,k},
\end{align}
then the inner product can be expanded as
\begin{align}
    \mathcal{I}(N,M\mid \mathbf{t}^+,\mathbf{t}^-) = \sum_{\lambda \subseteq [N,M]} s_\lambda(\mathbf{x}) \, s_\lambda(\mathbf{y}) .
\end{align}
Here \(s_\lambda(\mathbf{x})\) denotes the Schur polynomial in the variables \(x_i\), and the sum runs over Young diagrams \(\lambda\) contained in the \(N\times M\) rectangle.  In the limit \(N,M\to\infty\) this becomes the standard Toda tau function.

These functions are all Hirota tau functions because they admit a free fermion representation obtained by inserting the projection operators~\eqref{projectionoperator}, which are group-like elements:
\begin{align}
    \bra{N+1} \, e^{\mathfrak{J}_+(\mathbf{t}^+)} \, \mathrm{P}_{M+1}^- \, \mathrm{P'}_1^+ \, e^{\mathfrak{J}_-(\mathbf{t}^-)} \, \ket{N+1}.
\end{align}
This expression is precisely a Hirota tau function in the free fermion formalism.  When \(M\to\infty\) it reduces to the unitary matrix integral~\eqref{Todamatrixmodel}.

\subsubsection{Mapping between the bosonic chain and free fermions}

There exists a one-to-one mapping from the bosonic Fock space to the free-fermion Hilbert space.  Denoting the bosonic and fermionic basis states by
\begin{align}
    \begin{split}
        \ket{\mathbf{n}}_B &= \ket{n_1}_B \otimes \ket{n_2}_B \otimes \cdots \otimes \ket{n_L}_B, \qquad \sum_{j=1}^L n_j = N,\\
        \ket{\mathbf{m}}_F &= \ket{m_1}_F \otimes \ket{m_2}_F \otimes \cdots \otimes \ket{m_M}_F, \qquad \sum_{j=1}^M m_j = N,
    \end{split}
\end{align}
we can define a map \(P\) from the bosonic chain to the fermionic chain by
\begin{align}
    P \ket{n}_B = \underbrace{\ket{1}_F \otimes \cdots \otimes \ket{1}_F}_{\text{\(n\) times}} \otimes \ket{0}_F
    \quad \text{or, equivalently,} \quad
    \underbrace{\ket{\downarrow}_F \otimes \cdots \otimes \ket{\downarrow}_F}_{\text{\(n\) times}} \otimes \ket{\uparrow}_F .
\end{align}
The map is extended to the full bosonic Hilbert space site by site:
\begin{align}
    P \ket{\mathbf{n}}_B = \bigotimes_{i=1}^L P \ket{n_i}_B .
\end{align}
Under this correspondence, the Hamiltonian of the phase model maps to that of the so-called modified quantum XX spin chain~\cite{Pozsgay_2016}.

\subsubsection{Toda tau function in spin chain}

In the free-fermion construction one can also introduce the coherent states
\begin{align}
    \ket{\mathbf{t}} = \sum_{\mu} s_\mu(\mathbf{t}) \ket{\mu},
\end{align}
where \(s_\mu(\mathbf{t})\) stands for the Schur function of the variables \(x_i\) encoded by the Miwa coordinates \(t_k = \frac{1}{k}\sum_i x_i^k\).  In terms of these states, the matrix integral~\eqref{Todamatrixmodel} with \(N\to\infty\) becomes
\begin{align}
    \tilde{Z}_\infty(\mathbf{t}^+,\mathbf{t}^-) = \braket{\mathbf{t}^+}{\mathbf{t}^-} .
\end{align}
For finite \(N\) we must insert projection operators that restrict the sum to Young diagrams of length \(\ell(\lambda) \le N\):
\begin{align}
    \tilde{Z}_N(\mathbf{t}^+,\mathbf{t}^-) = \bra{\mathbf{t}^+} \prod_{r < -N} (1 - N_r) \ket{\mathbf{t}^-},
\end{align}
where \(N_r\) is the fermion number operator at site \(r\).

In the spin-chain realisation, working at finite \(N\) amounts to choosing the \(N\)-magnon vacuum \(\ket{\emptyset,N}\).  Using the non-commutative Cauchy identity~\eqref{NC-Cauchy} and the vacuum property~\eqref{svacyoung}, one finds
\begin{align}
    \prod_{i=1}^\infty \hat{H}(x_i) \ket{\emptyset,N}
    = \sum_{\lambda} s_\lambda(\mathbf{x}) \, \hat{s}_\lambda \ket{\emptyset,N}
    = \sum_{\ell(\lambda)\le N} s_\lambda(\mathbf{x}) \ket{\lambda},
\end{align}
with
\begin{align}
    \hat{H}(x) = \exp\!\Biggl( \sum_{k=1}^{\infty} \frac{x^k}{k} \, \hat{p}_k \Biggr).
\end{align}
Consequently, the finite \(N\) Toda tau function is expressed as the spin chain vacuum correlator
\begin{align}
    \tilde{Z}_N(\mathbf{t}^+,\mathbf{t}^-) = \bra{\emptyset,N} \,
    \exp\!\Biggl( \sum_{k=1}^{\infty} \frac{t_k^-}{k} \, \hat{p}_k^\dagger \Biggr) \,
    \exp\!\Biggl( \sum_{k=1}^{\infty} \frac{t_k^+}{k} \, \hat{p}_k \Biggr) \,
    \ket{\emptyset,N},
\end{align}
where the Miwa variables are related to the power sums by \(t_k^+ = p_k(\mathbf{x})\) and \(t_k^- = p_k(\mathbf{y})\).

Finally, KP and MKP tau functions, which arise as special limits of Toda tau functions, can also be rewritten as correlation functions between two \(N\) magnon vacua in the spin-chain Hilbert space.

\section{Conclusion}\label{Conclusion}

In this paper, we have established a precise correspondence between a broad class of unitary matrix models and vacuum correlation functions of quantized symmetric functions in the \(N\) magnon sector of quantum spin chains.  Starting from the free-fermion construction of the spin-chain Hilbert space and the quantization formalism of symmetric functions developed by Fomin and Greene, we have shown that the partition function of any unitary matrix model whose integrand admits a factorized expansion in power-sum symmetric functions can be recast as
\[
Z = \bra{\emptyset,N} \, \normord\hat{f}\normord \, \ket{\emptyset,N},
\]
where \(\hat{f}\) is built from the quantized power-sum operators \(\hat{\powsym}_k\) and \(\hat{\powsym}_k^\dagger\), which are identified with the conserved currents of the free-fermion system.  The central technical ingredient is the observation that the orthogonality of the Schur functions under the unitary group integral translates directly into the inner product of the quantized Schur states in the spin-chain Hilbert space.

We have illustrated the general framework with several explicit examples: GWW model, superconformal indices of $\mathcal{N}=4$ SYMs and Toda tau functions. These examples demonstrate that the mapping between unitary matrix integrals and spin chain correlators is both general and flexible, encompassing models that range from simple one-matrix integrals to supersymmetric gauge theory indices and integrable hierarchies.

\subsection*{Outlook}

The results presented in this paper open several promising directions for future investigation:

\begin{enumerate}
    \item \textbf{Quantum/Classical duality in integrable systems.}
    The deep connections between classical symmetric functions and superconformal indices~\cite{Hatsuda:2025mvj,Ren:2025tvx,Ren:2026dlj,Gadde:2011uv,Gaiotto:2012xa,Razamat:2013qfa,Rastelli:2014jja} were mentioned in the introduction.  In this paper we have shown that, by quantizing symmetric functions, the superconformal index can be related to quantum integrable systems.  This bridge may shed light on the broader Quantum/Classical duality in integrable systems.

    \item \textbf{Phase transitions and large \(N\) limits.}
    The representation of matrix models as spin chain correlators makes their large \(N\) behaviour amenable to techniques from the thermodynamic Bethe ansatz and quantum quench dynamics.  It would be particularly interesting to study the third-order Gross--Witten phase transition and the Hawking--Page transition of the superconformal index directly in the spin chain language, possibly relating them to dynamical critical phenomena in the XXZ chain.

    \item \textbf{Quantum algorithms.}
    The MPS--MPO construction of quantized symmetric functions and the quantum circuits for character computations discussed in the introduction suggest that the evaluation of unitary matrix integrals could be reformulated as a problem in quantum computation.  Whether the vacuum correlators obtained here can be efficiently simulated on a quantum computer, and what this implies for the computational complexity of supersymmetric indices and topological string amplitudes, remain open questions.

    \item \textbf{Further integrable structures.}
    The quantized power-sum operators satisfy a closed current algebra, and their exponentials are vertex operators.  This hints at an underlying quantum group structure, which may be related to the Yangian or quantum affine algebras that appear in the algebraic Bethe ansatz.  Clarifying this algebraic structure could reveal new conservation laws and exact relations among the matrix model observables.

    \item \textbf{Other matrix ensembles.}
    While this paper focused on unitary matrix models, the method should extend to orthogonal and symplectic ensembles, where the Schur basis is replaced by the corresponding orthogonal or symplectic characters.  The results of Section~\ref{superconformal indices of N=4 SYMs} for the \(Sp(2N)\) and \(SO(N)\) indices are a first step in this direction, and a systematic treatment would be valuable.

    \item \textbf{Generalizations to other families of symmetric functions.}
    In this paper we have considered only the quantization of the \(e, h, \powsym, s\) families of symmetric functions.  Extending the procedure to Macdonald polynomials, elliptic Macdonald polynomials and Koornwinder polynomials, which depend on additional parameters, would be a natural and interesting next step.
\end{enumerate}

We hope that the formalism developed here will provide a useful bridge between matrix models, quantum integrable systems, and symmetric functions, and will stimulate further developments at the interfaces of these and related areas.

\section*{Acknowledgements}
I thanks Minxin Huang and Xin Wang for helpful discussions. I am supported by the National Natural Science Foundation of China Grants No.12325502 and No.12247103.

\appendix
\section{Appendix}\label{appendixA}
\subsection{Proof of~\eqref{svacyoung}}\label{Proof ofeqrefsvacyoung}

We briefly sketch the proof of \eqref{svacyoung}.  It is convenient to describe the Young diagram \(\lambda\) by its frequency representation \(\lambda = (1^{n_1} 2^{n_2} \cdots M^{n_M})\), i.e.\ the diagram contains \(n_1\) rows of length \(1\), \(n_2\) rows of length \(2\), and so on, up to length \(M\).  The total number of rows is \(\ell(\lambda) = n_1 + n_2 + \cdots + n_M\).  The \(N\)-magnon vacuum is \(\ket{\emptyset,N} = \ket{\downarrow^N \uparrow\uparrow\cdots}\).

According to the definition of the quantized Schur function \eqref{defShat} and the identification \eqref{defxi}, each column of a semistandard tableau contributes a product of shift operators \(\hat x_i\).  Because the columns act on the state sequentially from left to right, the first column must be chosen so that its action is non-vanishing.  For a column of height \(\ell(\lambda)\) the only non-zero contribution comes from the tableau whose entries increase by one step downwards and occupy the lowest possible positions, namely
\begin{align*}
   \Yboxdimx{50pt}
   \Yboxdimy{15pt}
   \young({\scriptstyle N-\ell{(\lambda)}+1},\vdots,{N-1},N) .
\end{align*}
This column moves the last \(\ell(\lambda)\) down spins one step to the right.  

Proceeding to the second column, its height is \(\ell(\lambda) - n_1\) (since the first \(n_1\) rows have length \(1\) and do not extend beyond one column).  Again, to avoid vanishing it must be placed as low as possible, immediately to the right of the first column.  Thus the first two columns take the form
\begin{align*}
   \Yboxdimx{70pt}
   \Yboxdimy{18pt}
   \young({\scriptstyle N-\ell{(\lambda)}+1}{\scriptstyle N+1-\ell{(\lambda)}+n_1+1},\vdots \vdots,{\scriptstyle N-n_1}{N},{N-n_1+1}{N+1},\vdots,{N-1},{N}) .
\end{align*}
The second column then shifts the last \(\ell(\lambda)-n_1\) of the already displaced fermions one further step to the right.  

Continuing in this way, the \(j\)-th column acts on the last \(\ell(\lambda) - (n_1 + \cdots + n_{j-1})\) fermions.  After all columns have been applied, the original block of \(N\) down spins has been split into groups separated by up spins exactly according to the row lengths of \(\lambda\).  The final state reads
\begin{align}
    \ket{\lambda,N} = \ket{\downarrow^{N-M} \uparrow^{n_M} \downarrow \uparrow^{n_{M-1}} \downarrow \uparrow^{n_{M-2}} \cdots \downarrow \uparrow^{n_1} \downarrow \uparrow\cdots},
\end{align}
which is the basis vector associated with the Young diagram \(\lambda\) in the \(N\)-magnon sector.  Hence \(\hat s_\lambda \ket{\emptyset,N} = \ket{\lambda,N}\), completing the proof.

\subsection{Proof of~\eqref{quantizedps} and~\eqref{quantizedpmunu}}\label{Proof ofeqrefquantizedps}

We now give a detailed derivation of the action of the quantized power-sum operators \(\hat p_r\) on the basis states \(\ket{\lambda}\), showing that it reproduces the classical Murnaghan--Nakayama rule for the skew characters of the symmetric group.

Recall the frequency representation of a Young diagram \(\lambda = (1^{n_1} 2^{n_2} \cdots M^{n_M})\), which means \(\lambda\) contains \(n_1\) rows of length \(1\), \(n_2\) rows of length \(2\), and so on.  The corresponding state in the \(N\)-magnon sector is
\begin{align*}
    \ket{\lambda} = \ket{\downarrow^{n_0} \uparrow \downarrow^{n_1} \uparrow \downarrow^{n_2} \uparrow \cdots \uparrow \downarrow^{n_M} \uparrow\cdots},
\end{align*}
where \(n_0 = N - \sum_{i=1}^M n_i\) is the number of trailing down spins before the first up spin.

The operator \(\hat p_r = \sum_{i\ge 1} c_{i+r}^\dagger c_i\) acts on a configuration by removing a fermion (a down spin) at site \(i\) and creating one at site \(i+r\).  When applied to \(\ket{\lambda}\), the only non-vanishing contributions come from terms where the annihilated fermion sits inside one of the blocks \(\downarrow^{n_i}\) and the created fermion lands in a region that was originally empty (an up spin) with an overall factor $\pm1$, we first ignore this sign and determine it at the end. This process changes the shape of the Young diagram.  Let us examine the effect in detail.

Consider a term that annihilates a down spin inside the \(n_i\)-th block and creates one inside the \((i+m)\)-th up-spin region (i.e.\ after the block \(\downarrow^{n_{i+m}}\)).  The state \(\ket{\lambda}\) can be written as
\begin{align}\label{lambda}
    \ket{\lambda} = \ket{ \cdots \uparrow \downarrow^{n_i} \uparrow \downarrow^{n_{i+1}} \cdots \downarrow^{n_{i+m-1}} \uparrow \downarrow^{n_{i+m}} \cdots }.
\end{align}
The action of \(c_{i+r}^\dagger c_i\) changes it to
\begin{align}\label{lambda'}
    \ket{\lambda'} = \ket{ \cdots \uparrow \downarrow^{n'_i} \uparrow \downarrow^{n'_{i+1}} \cdots \downarrow^{n'_{i+m-1}} \uparrow \downarrow^{n'_{i+m}} \cdots },
\end{align}
where the configurations outside the displayed segment are identical.  A simple counting shows that the new multiplicities \(n'_j\) are related to the old ones by
\begin{align}\label{nandn'}
    n'_i + n'_{i+1} + 1 = n_i, \qquad
    n'_{i+m} = n_{i+m-1} + n_{i+m} + 1, \qquad
    n'_{i+k} = n_{i+k-1} \quad (1 < k < m).
\end{align}
These relations follow because one down spin is removed from the first block and one is added after the \(m\)-th block, effectively shifting one fermion across \(m\) blocks of down spins and \(m\) intervening up spins.

\begin{figure}[t]
\begin{center}
\includegraphics{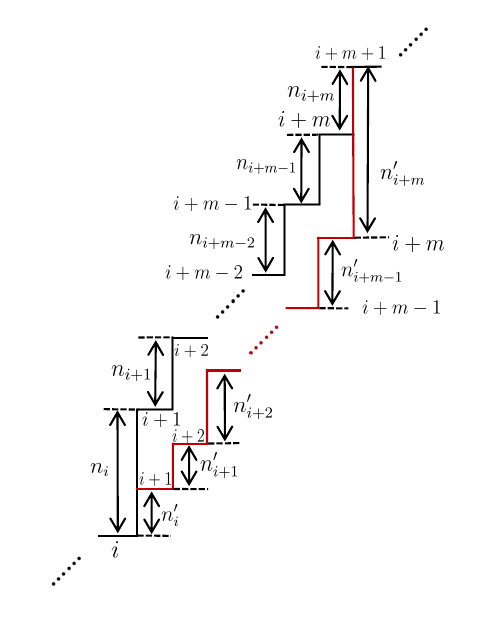}
\caption{$\ket{\lambda}$ and $\ket{\lambda'}$. 
The black solid line is the boundary of Young diagram $\lambda$, and the red solid line is the boundary of Young diagram $\lambda'$.}
\label{pscharacter}
\end{center}
\end{figure}

The total number of boxes in the Young diagram associated with \(\ket{\lambda}\) is
\begin{align*}
    |\lambda| = \sum_{k\ge 1} k \, n_k.
\end{align*}
Using \eqref{nandn'}, a short calculation gives
\begin{align*}
    |\lambda'| - |\lambda| = r.
\end{align*}
The relations \eqref{nandn'} imply that \(\lambda'/\lambda\) is a border strip. Since the number of boxes changes by \(r\), this border strip has size \(r\). Conversely, any \(\lambda'\) for which \(\lambda'/\lambda\) is a border strip of size \(r\) can be obtained by shifting a down spin \(r\) sites to the right. Therefore
\(\hat{p}_r=\sum_{i=1}^\infty c^\dagger_{i+r}c_i\) sums over all such shifts and generates precisely the diagrams \(\lambda'\) with \(\lambda'/\lambda\) a border strip of size \(r\). The same argument applies to a general partition \(\lambda=(0^{n_0},1^{n_1},\cdots,M^{n_M})\).
Thus \(\lambda'\) is obtained from \(\lambda\) by adding a border strip of size \(r\). The geometric correspondence is illustrated in Figure~\ref{pscharacter}.
Using the isomorphism between the spin-chain basis and the free-fermion construction, we denote a down spin state at the \(i\)-th site by
 \begin{align}
     \ket{\downarrow}_i:=c_i\ket{\uparrow}_i
 \end{align}
The fermionic operators satisfy the canonical anticommutation relations. Acting with \(\hat p_r\) on \(\lambda\) produces all \(\lambda'\) such that \(\lambda'/\lambda\) is a border strip of size \(r\), with coefficients \(c_\lambda(\lambda')=\pm1\). To determine the sign, recall that \(\hat p_r = \sum_{k\ge 1} c_{k+r}^\dagger c_k\). When this operator acts on a state \(\ket{\lambda}\), the term with a given \(k\) first annihilates a fermion at site \(k\), which must be occupied by a down spin, and then creates one at site \(k+r\), which must be an up spin. The fermionic anticommutation relations imply that moving a down spin past another down spin produces a minus sign. Consequently, as the fermion travels from site \(k\) to site \(k+r\), it picks up a factor \(-1\) for every down spin between these two positions. The number of such down spins is the height \(\operatorname{ht}(\lambda'/\lambda)\) of the border strip. Therefore each border strip contributes the factor
\begin{align}
    (-1)^{\text{ht}(\lambda'/\lambda)}\,,
\end{align}
so
\begin{align}\label{quantizedpactonlambda}
\hat{p}_r\ket{\lambda}=\sum_{\lambda'}(-1)^{\text{ht}(\lambda'/\lambda)}\ket{\lambda'}\,,
\end{align}
where summation is over all $\lambda'$ such that $\lambda'/\lambda$ is a border strip of size $r$.

We now verify that the action of the quantized power-sum operators reproduces the Murnaghan--Nakayama rule for skew characters, thereby proving \eqref{quantizedps}.  Using the definition \(\hat p_\mu = \hat p_{\mu_1} \cdots \hat p_{\mu_{\ell(\mu)}}\) and the basic action \eqref{quantizedpactonlambda}, we can build the character formula recursively.  

Now apply a product \(\hat p_{\mu_1} \cdots \hat p_{\mu_n}\) to the vacuum \(\ket{\emptyset}\).  By successive insertions of complete sets of states, the coefficient of \(\ket{\lambda}\) in the resulting vector is
\begin{align}
    \chi^\lambda(\mu_1,\dots,\mu_n) = \sum_{\nu^{(1)},\dots,\nu^{(n-1)}} 
\prod_{j=1}^{n} (-1)^{\operatorname{ht}(\nu^{(j-1)}/\nu^{(j)})}\,,
\end{align}
where \(\nu^{(0)} = \lambda\), \(\nu^{(n)} = \emptyset\), and each skew diagram \(\nu^{(j-1)}/\nu^{(j)}\) is a border strip of size \(\mu_{n-j+1}\).  This is precisely the recursive combinatorial definition of the irreducible character of the symmetric group, known as the Murnaghan--Nakayama rule.  

For the skew version, one simply replaces the vacuum by an arbitrary state \(\ket{\nu}\) and obtains
\begin{align}
  \hat p_\mu \ket{\nu} = \sum_{\lambda} \chi^{\lambda/\nu}(\mu) \ket{\lambda}\,,  
\end{align}
which is exactly the operator form of \eqref{quantizedps}.  Thus the quantized power-sum operators act as the skew characters on the spin-chain Hilbert space.

To see that this guarantees the full ring structure, recall that for classical symmetric functions one has the identity
\begin{align}
    p_\mu s_\lambda = \sum_{\nu} \chi^{\nu}(\mu) s_\nu s_\lambda 
= \sum_{\nu,\gamma} \chi^{\nu}(\mu) C_{\nu\lambda}^{\gamma} s_\gamma 
= \sum_{\gamma} \chi^{\gamma/\lambda}(\mu) s_\gamma,
\end{align}
so that \(\sum_{\nu} \chi^{\nu}(\mu) C_{\nu\lambda}^{\gamma} = \chi^{\gamma/\lambda}(\mu)\).  In the quantized theory, because the Schur operators multiply with the same Littlewood--Richardson coefficients \(C_{\mu\nu}^{\lambda}\) (see \eqref{hatSSS}), the same combinatorial identity holds, and we obtain
\begin{align}
    \hat p_\mu \,\hat s_\nu = \sum_{\lambda} \chi^{\lambda/\nu}(\mu) \,\hat s_\lambda\, .
\end{align}

Applying this operator equation to the vacuum and comparing with the direct action of \(\hat p_\mu\) on the states proves the consistency of the quantized ring for the full \((e,h,p,s)\) basis.  Hence, on the \(N\) magnon Hilbert space, the quantized power-sum operators satisfy
\begin{align}
    \hat p_\mu = \sum_{\lambda} \chi^{\lambda/\mu} \,\hat s_\lambda\, .
\end{align}


\bibliographystyle{JHEP} 
\bibliography{References}
\end{document}